\documentclass[amsmath,amssymb,superscriptaddress,nobalancelastpage,aps,twocolumn,showpacs]{revtex4-1}

\usepackage{hyperref}
\hypersetup{colorlinks,linkcolor=blue,urlcolor=blue,citecolor=blue}

\usepackage{graphicx}
\usepackage{dcolumn}
\usepackage{bm}

\usepackage{color}
\usepackage{ulem}

\begin{document}

\title{Superconducting gap and pseudogap in the surface states of the iron-based
superconductor PrFeAsO$_{1-y}$ studied by angle-resolved photoemission spectroscopy}

\author{K. Hagiwara}
\email{k.hagiwara@fz-juelich.de}
\affiliation{Department of Physics, University of Tokyo, Bunkyo-ku, Tokyo 113-0033, Japan}

\author{M. Ishikado}
\affiliation{Comprehensive Research Organization for Science and Society (CROSS), Tokai, Ibaraki 319-1106, Japan}

\author{M. Horio}
\affiliation{Department of Physics, University of Tokyo, Bunkyo-ku, Tokyo 113-0033, Japan}

\author{K. Koshiishi}
\affiliation{Department of Physics, University of Tokyo, Bunkyo-ku, Tokyo 113-0033, Japan}

\author{S. Nakata}
\affiliation{Department of Physics, University of Tokyo, Bunkyo-ku, Tokyo 113-0033, Japan}

\author{S. Ideta}
\affiliation{UVSOR Synchrotron Facility, Institute for Molecular Science, Okazaki, Aichi, 444-8585, Japan}

\author{K. Tanaka}
\affiliation{UVSOR Synchrotron Facility, Institute for Molecular Science, Okazaki, Aichi, 444-8585, Japan}

\author{K. Horiba}
\affiliation{Photon Factory, High Energy Accelerator Research Organization (KEK), Tsukuba, Ibaraki 305-0801, Japan}

\author{K. Ono}
\affiliation{Photon Factory, High Energy Accelerator Research Organization (KEK), Tsukuba, Ibaraki 305-0801, Japan}

\author{H. Kumigashira}
\affiliation{Photon Factory, High Energy Accelerator Research Organization (KEK), Tsukuba, Ibaraki 305-0801, Japan}

\author{T. Yoshida}
\affiliation{Department of Interdisciplinary Environment, Kyoto University, Sakyo-ku, Kyoto 606-8501, Japan}

\author{S. Ishida}
\affiliation{National Institute of Advanced Science and Technology (AIST), Tsukuba, Ibaraki 305-8568, Japan}

\author{H. Eisaki}
\affiliation{National Institute of Advanced Science and Technology (AIST), Tsukuba, Ibaraki 305-8568, Japan}

\author{S. Shamoto}
\affiliation{Comprehensive Research Organization for Science and Society (CROSS), Tokai, Ibaraki 319-1106, Japan}
\affiliation{Japan Atomic Energy Agency (JAEA) , Tokai, Ibaraki 319-1195, Japan}
\affiliation{Department of Physics, National Cheng Kung University, Tainan 70101, Taiwan}
\affiliation{Meson Science Laboratory, RIKEN, Wako, Saitama 351-0198, Japan}

\author{A. Fujimori}
\affiliation{Department of Physics, University of Tokyo, Bunkyo-ku, Tokyo 113-0033, Japan}
\affiliation{Department of Applied Physics, Waseda University, Shinjuku-ku, Tokyo 169-8555, Japan}

\date{\today}

\begin{abstract}

In order to study the possible superconductivity at the polar surfaces of 1111-type iron-based superconductors, which is doped with a large amount of holes in spite of the electron doping in bulk materials, we have performed angle-resolved photoemission spectroscopy (ARPES) studies on superconducting PrFeAsO$_{1-y}$~crystals. We have indeed observed the opening of a superconducting gap on surface-derived hole pockets as well as on a bulk-derived hole pocket. 
The superconducting gap is found to open on the surface-derived hole pockets below the bulk $T_\text{c}$, which suggests that the surface superconductivity is possibly induced by proximity effect from the bulk.
We have also observed the opening of a large pseudogap on the surface-derived hole pockets, which is similar to the pseudogap in 122-type bulk superconductors doped with a smaller amount of holes. This suggests that the opening of a large pseudogap is a characteristic property of hole-doped iron-based superconductors.

\end{abstract}

\maketitle

\section{\label{sec:level1}Introduction}

The discovery of iron-based superconductors has attracted tremendous attention in materials science during the last decade \cite{Kamihara2008}. 
$Ln$FeAsO ($Ln$=Lanthanoide) compounds, so-called 1111-type superconductors, were first discovered and have the highest superconducting transition temperature ($T_\text{c}$ $\sim$ 58 K) among the iron-based superconductors. 
In the 1111-type iron pnictides, superconductivity is realized by electron doping through substitution of F for O \cite{Kamihara2008}, H for O \cite{Iimura2012}, Co for Fe \cite{Wang2009}, or introducing oxygen deficiencies \cite{Lee2008}.
According to angle-resolved photoemission spectroscopy (ARPES) studies, however, cleaved surfaces of the 1111 compounds exhibit heavily hole-doped electronic structures. The hole doping at the surface has been attributed to the surface polarity of the 1111 compounds, which are inevitable due to their crystal structures and cleavage planes \cite{Lu2008, Nishi2011, Zhang2016}. 
All the observed Fermi surfaces of the 1111 compounds were found to be $k_{z}$ independent, suggesting the two-dimensional nature of the observed electronic structure, which makes it difficult to distinguish between bulk- and surface-derived Fermi surfaces \cite{Liu2010}.
Charnukha $et~al$. \cite{Charnukha2015} separated ARPES spectra into bulk and surface states using first-principles calculation on bulk crystals.
They observed three hole pockets centered at the zone center, namely, "Outer", "Middle", and "Inner" hole pockets, and assigned the Outer and Middle hole pockets to the surface bands, and the Inner hole pocket to bulk bands.

According to ARPES studies, the heavily hole-doped, cleaved surfaces of 1111 compounds are superconducting \cite{Charnukha2015, Charnukha2016}.
The momentum dependence of the superconducting gap of NdFeAsO$_{0.9}$F$_{0.1}$ ($T_\text{c}$ $\sim$ 53 K) indicates the opening of an isotropic or weakly anisotropic s-wave gap on the hole Fermi surface \cite{Kondo2008}.
Other angle-integrated photoemission studies indicate a large pseudogap gap opening above $T_\text{c}$ \cite{Sato2008, Sato2008a, Nakamura2017}. Thus, the surface of the 1111-type superconductors is unique for its heavy hole doping but whether the superconductivity is originated from the surface or bulk electronic states is not yet clear. Therefore, the relationship between the surface electronic states, the possibility of superconductivity at the surface, and the origin of the pseudogap in the 1111-type superconductors remains to be investigated.

In this work, we have performed ARPES studies on cleaved single crystals of the 1111-type iron-based superconductor PrFeAsO$_{1-y}$, and have measured the temperature dependence of energy gaps in order to study the possible superconductivity at the surface of the 1111 system.
We have observed a superconducting gap and a pseudogap opening in the hole-doped surface states. We discuss our results in comparison with the superconducting gap and the pseudogap of the hole-doped 122-type superconductors.

\section{Experiment}
High-quality single crystals of the electron-doped superconductor PrFeAsO$_{1-y}$ ($T_\text{c}$ $\sim$ 16 and 33 K), which belongs to the 1111 family with oxygen deficiencies, were synthesized under high pressure as described in Ref. \cite{Ishikado2009, ishikado2010growth}.
ARPES measurements were performed at BL7U of UVSOR using linearly polarized light with the photon energy of 22.5 eV. 
An MBS A1 electron analyzer was used. 
The total energy resolution of the ARPES measurement was $\sim$ 10 meV.
The crystals were cleaved $in~situ$ at $T$ $\sim$ 12 K and measured in an ultrahigh vacuum of $\sim$ 10$^{-10}$ Torr.

\section{Result and discussion}

\begin{figure}[b]
\includegraphics[clip,width=8.5cm]{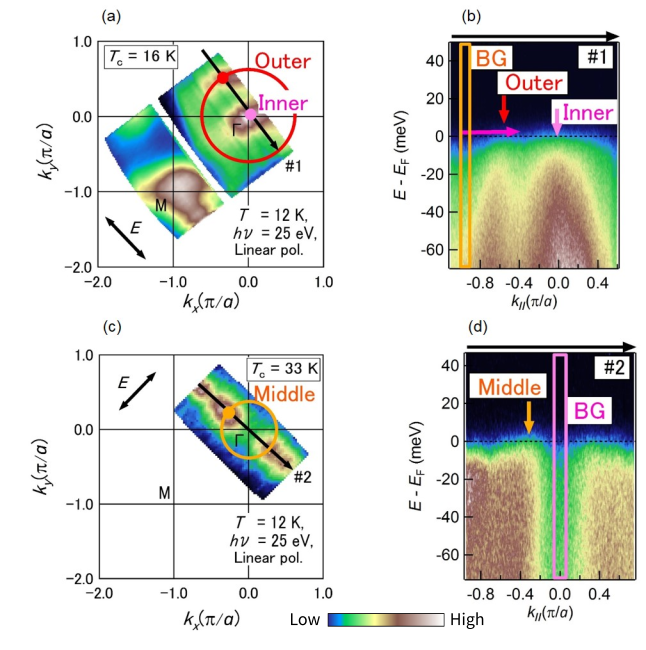}
\caption{
Fermi-surface mapping and band dispersions measured using linearly polarized light. (a) Fermi-surface mapping of PrFeAsO$_{1-y}$ ($T_\text{c}$ $\sim$ 16 K) measured using $p$ polarized light . (b) Energy($E$)-inplane-momentum($k_{\parallel}$) plot along cut \#1 indicated in panel (a). (c), (d) The same as (a) and (b) but of PrFeAsO$_{1-y}$ ($T_\text{c}$ $\sim$ 33 K) using $s$ polarized light. Outer and Inner hole pockets appear for $p$ polarized light, while Middle hole pocket appears for $s$ polarized light. Area enclosed by an orange square in panel (b) and by a pink square in panel (d) indicate background (BG) which is used for the analysis.
}\label{fig:1}
\end{figure}

Figure \ref{fig:1} shows Fermi-surface mapping and band dispersions of PrFeAsO$_{1-y}$ measured  using linearly polarized light. 
As shown in Figs. \ref{fig:1} (a) and (b) taken using $p$ polarized light for PrFeAsO$_{1-y}$ with $T_\text{c}$ $\sim$ 16 K, one can observe a large circular hole pocket and a small hole pocket around the zone center, referred to as "Outer" and "Inner" hole pockets, respectively.
In contrast, Figs. \ref{fig:1} (c) and (d) taken using $s$ polarized light for PrFeAsO$_{1-y}$ with $T_\text{c}$ $\sim$ 33 K show a circular hole pocket of the intermediate size,  referred to as a "Middle" hole pocket.
These observations are consistent with the previous reports \cite{Charnukha2015, Nishi2011}. 

\begin{figure}[b]
\includegraphics[clip,width=7.5cm]{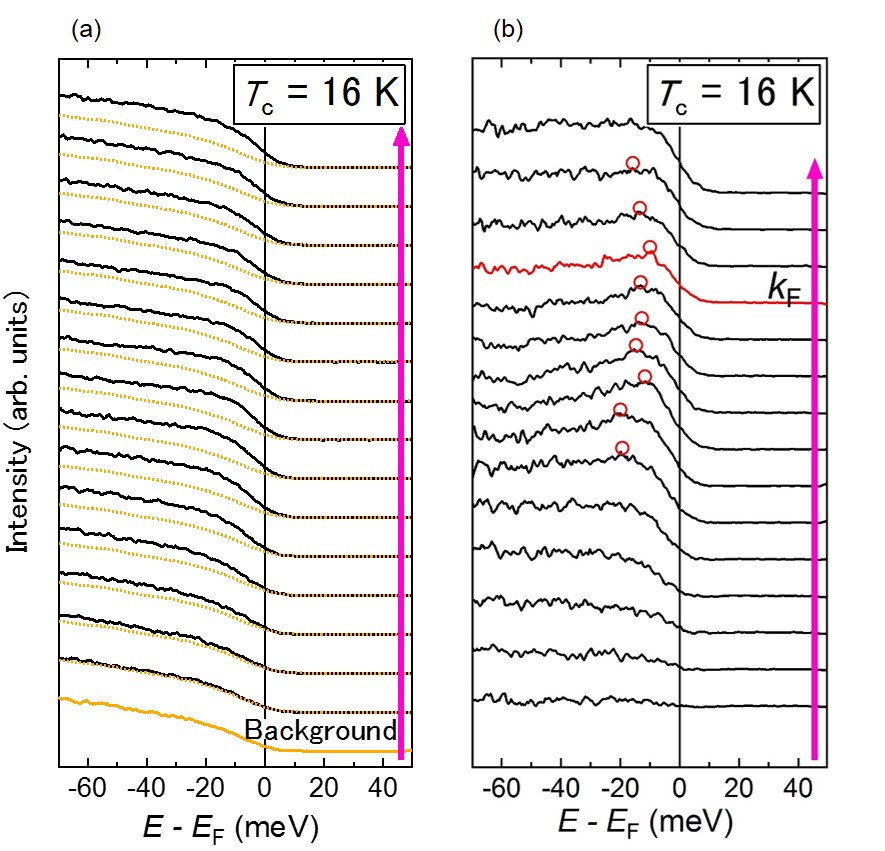}
\caption{
Analysis of the superconducting gaps on the Outer hole pocket. (a) Raw energy distribution curves (EDCs) along the pink arrow in Fig. \ref{fig:1} (b). (b) EDCs after background subtraction. Red open circles indicate the positions of the coherence peaks.
}\label{fig:2}
\end{figure}
    
We have analyzed energy gaps at various momenta on the hole pockets.
Figure \ref{fig:2} shows analysis of the energy gap on the Outer hole pocket, which is derived from the surface \cite{Charnukha2015}. As seen from the raw data of energy distribution curves (EDCs) in panel (a), one cannot clearly identify a coherence peak that indicates superconductivity. 
However, if one defines the background by an orange EDC in Fig. \ref{fig:2} (a), which is made from the averaged EDC in the orange square area in Fig. \ref{fig:1} (b), background subtraction from the raw EDCs yields coherence peaks indicated by red open circles in Fig. \ref{fig:2} (b) and the backbending of the peak dispersion which is reminiscent of the Bogoliubov quasi-particle expected for a superconducting gap opening.

\begin{figure}
\includegraphics[clip,width=8cm]{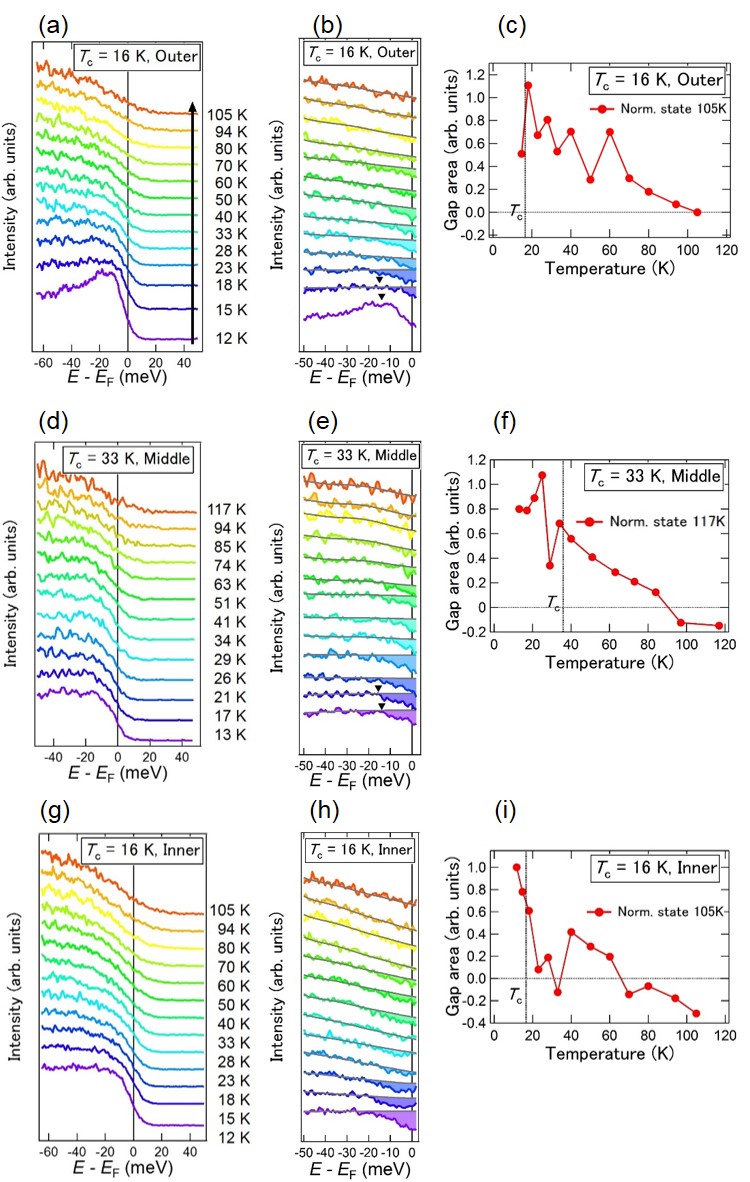}
\caption{
Temperature dependence of the EDCs and the energy gaps at the three hole pocket.
(a) Temperature dependence of the background-subtracted EDCs on the Outer hole pocket.
(b) Temperature dependence of the EDCs divided by the Fermi-Dirac function for the Outer hole pocket.
The normal-state EDC set at 117K/105K are overlaid on each EDC.
Black triangles indicate the gap size. Shaded areas indicate the gap area.
(c) Temperature dependence of the gap magnitude on the Outer Fermi surface.
(d, e, f) (g, h, i) The same as (a, b, c) but on the Middle and Inner hole pockets, respectively. (a, b, c) and (g, h, i) are obtained from the results for $p$ polarized light. (d, e, f) are obtained from the results for $s$ polarized light.
}\label{fig:3}
\end{figure}

\begin{figure}
\includegraphics[clip,width=8.5cm]{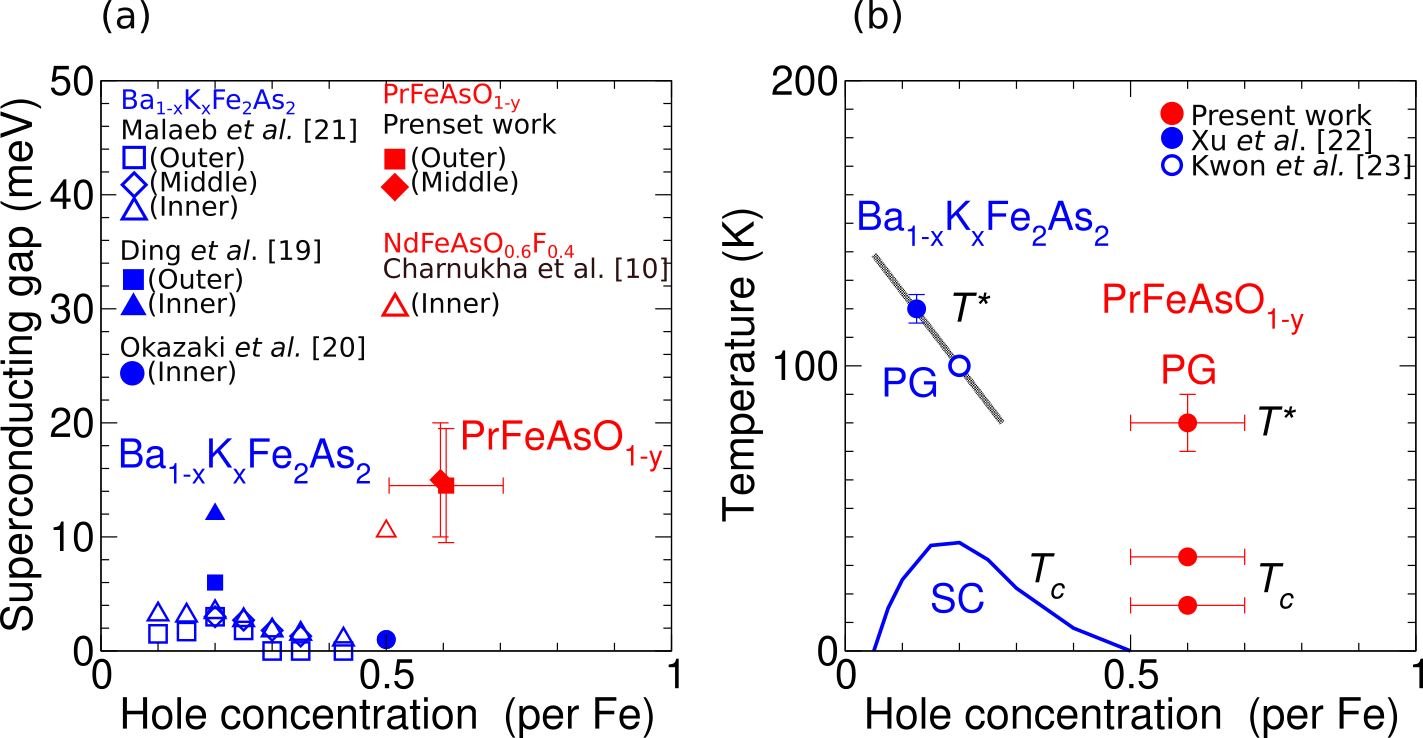}
\caption{
Summary of the superconducting gap and the pseudogap of the surface hole-doped PrFeAsO$_{1-y}$ (red markers) in comparison with those of the bulk hole-doped Ba$_{1-x}$K$_{x}$Fe$_{2}$As$_2$ (blue markers).
(a) Superconducting gap as a function of surface/bulk hole concentration.
(b) Pseudogap temperatures.
Squares, diamonds, and triangles indicate those of the Outer, Middle, and Inner hole pockets, respectively.
}\label{fig:4}
\end{figure}

Figures \ref{fig:3} (a), (b), and (c) show the temperature dependence of EDCs on the Outer hole pocket. The Fermi momentum $k_\text{F}$ has been determined by a Lorentzian fit to the momentum distribution curve (MDC) at $E_\text{F}$. In order to eliminate the influence of the Fermi Dirac (FD) function, we have devided the EDC by the FD function as shown in panel (b). One can see an energy gap opening and the emergence of a coherence peak at low temperatures.
In order to analyze the energy gaps, we first consider that the EDC taken at the highest temperature represents an EDC in the normal state (without a superconducting gap and a pseudogap), and we overlay it on other EDCs in panel (b).
Here, we have normalized  each EDC to the normal-state EDC in the high binding energy region ($E-E_\text{F}\leq-30$ meV).
As a measure of the gap magnitude, we have estimated the gap area enclosed by the FD-divided EDC and the normal-state EDC. 
In the case of EDCs below $T_\text{c}$, however, the normal-state EDCs cannot be used to fit the spectra outside the gap region because the low temperature spectrum are convex in that region. For those temperatures, therefore, we have fitted the spectrum outside the gap region to a convex hyperbola. The gap size has thus been estimated from the peak positions, as indicated by black triangles in panel (b).
Panel (c) shows the temperature dependence of the gap magnitude.
One finds that the energy gap remains open above $T_\text{c}$ and the coherence peaks are observed below $T_\text{c}$.
We have also made analysis by taking spectra at other temperatures as the normal-state EDC and found consistent behavior of the gap area as shown in Supplementary information.

In the Middle hole pocket, which also originates from the surface states \cite{Charnukha2015}, we have analyzed the data in the same way as those of the Outer hole pockets, as shown in panels (d), (e), and (f). In this case, we define the background spectrum as indicated by a pink square in Fig. \ref{fig:1} (d).
The result shown in panel (f) is similar to the result of the Outer hole pocket. One can observe coherence peaks below the bulk $T_\text{c}$, which suggests that the gap is a superconducting gap.
The energy gap remains open above $T_\text{c}$, which indicates a pseudogap, too.

Figures \ref{fig:3} (g), (h), and (i) show the results of the Inner hole pocket, which originates from the bulk \cite{Charnukha2015}. One can see that the energy gap suddenly decreases above $T_\text{c}$ and closes faster than the Outer and Middle hole pockets. We consider that the different behavior of the Inner hole pocket from the Outer and Middle ones may reflect the bulk origin of the superconductivity for the Inner hole pocket. 

Thus our results on the 1111 compound show a gap of 14.5 $\pm$ 5 and 15 $\pm$ 5 meV for the Outer and Middle hole pockets, respectively, which indicates nearly the same gap sizes for the different hole pockets. The present result of the gap size is consistent with the previous reports (a gap of 15 meV on the hole pocket for another 1111 compound NdFeAsO$_{0.9}$F$_{0.1}$ \cite{Kondo2008} and a gap of 10.5 meV on the Inner hole pocket for NdFeAsO$_{0.6}$F$_{0.4}$ \citep{Charnukha2016}) although whether the superconducting gap is of the bulk or surface origin has not been pursued in Ref. \cite{Kondo2008}.

Now, we would like to consider the origin of the superconducting gap in the Outer and Middle hole pockets originating from the surface.
The out-of-plane coherence length in the 1111-type superconductors is given by $\xi_{c}~\sim$ 4.5 \AA ~\cite{okazaki2009lower, van2010flux}.
Because this $\xi_{c}$ is only a little smaller than the distance $\sim$ 8 \AA  ~between the FeAs planes, it may be possible that the superconducting gap in the surface bands is induced by proximity effect from the bulk superconductivity \cite{Charnukha2015, Charnukha2016}.
(Note that the escape depth of photon electrons excited by VUV light is 5-10 \AA, which is comparable to the distance between the FeAs planes.)

In Fig. \ref{fig:4} (a), we compare the present results on the superconducting gaps at the hole-doped surfaces of the 1111 compounds with the bulk superconductivity in heavily hole-doped 122 compounds. The surfaces of the 1111 compounds are doped with 0.5-0.7 holes per Fe \cite{Nishi2011, Zhang2016} and the superconducting gaps are as large as $\sim$ 15 meV. 
On the other hand, the superconducting gaps of the hole-doped 122 compound Ba$_{1-x}$K$_{x}$Fe$_{2}$As$_2$ become as large as 6-12 meV for the hole concentration of 0.2 per Fe atom for $x=0.4$ \cite{Ding2008} but becomes as small as $\sim$1 meV for  $x=1$ (for 0.5 hole per Fe atom as in the surface of the 1111 compounds) \cite{okazaki2012octet}.
Malaeb $et~al.$ \cite{Malaeb2012} have shown that the superconducting gap in the inner hole pocket roughly scale with $T_\text{c}$.
Therefore, the surface superconductivity observed in the present work should arise from a mechanism distinct from the superconductivity of the hole-overdoped bulk compounds. 
The superconductivity of the surface-derived band is possibly caused by proximity effect of the bulk superconductivity. The present observation that the $T_\text{c}$ of the surface superconductivity, which has been estimated from the observed coherence peaks on the surface-derived bands, is almost the same as the  $T_\text{c}$ of bulk superconductivity is consistent with the proximity-induced superconductivity.

Our results also show that a pseudogap is clearly observed below the pseudogap temperature $T^{*} \sim 80$ K in the Outer and Middle hole pockets, but not clearly in the Inner one, which was not able to conclude by angle-integrated photoemission studies \cite{Sato2008, Sato2008a, Nakamura2017}.
A pseudogap was observed in bulk  Ba$_{1-x}$K$_{x}$Fe$_{2}$As$_2$ up to  $T^{*} \sim 120$ K for $x=0.25$ \cite{Xu2011} and $T^{*} \sim 100$ K for $x=0.4$ \cite{kwon2012evidence}.
Considering the previous reports on the hole-doped 122 compound summarized in Fig. \ref{fig:4} (b), we conclude that the large pseudogap is common to hole-doped iron-superconductors. 
The similar pseudogaps observed in angle-integrated photoemission spectroscopy (AIPES) spectra of 1111 compounds \cite{Sato2008, Sato2008a, Nakamura2017} also open below $T^{*} \sim 100$ K or 150 K. Although the surface doping levels could not be estimated from the AIPES studies, it is likely that the surfaces were hole-doped as in the present case as well as in the previous ARPES studies on the 1111 compounds \cite{Lu2008, Nishi2011, Zhang2016, Liu2010, Charnukha2015, Charnukha2016, Kondo2008}.
It should be noted that the Inner hole pocket arising from the bulk 1111 compound, which is moderately electron-doped, does not show a clear pseudogap.

\section{Conclusion}
We have performed ARPES studies of the 1111-type iron-based superconductor PrFeAsO$_{1-y}$, and have measured the temperature dependence of the superconducting gap and the pseudogap.
In agreement with the previous reports, heavy hole-doping due to the surfaces polarity has been observed.
In the Outer and Middle hole pockets originating from the surface states, we have observed a coherence peak below the bulk $T_\text{c}$, which implies that the surface superconductivity is caused by proximity effect from the bulk 1111 superconductor.
The energy gap of the surface-derived hole pocket remains open above $T_\text{c}$, indicating a pseudogap opening.
In the Inner hole pocket originating from the bulk, we have observed that the energy gap suddenly decreases above $T_\text{c}$ and pseudogap behavior is less pronounced than the surface hole pockets, consistent with the behavior of the electron-doped bulk 1111 superconductors.  
Comparing the present results on the surface of the 1111 superconductor with those of hole-doped 122 compounds, we find that the superconducting gap is much larger at the surface of the 1111 superconductor than the hole-overdoped bulk 122 compounds. On the other hand, the pseudogap is equally large, suggesting that the pseudogap is common to hole-doped iron-based superconductors, irrespective of the bulk or surface origin of superconductivity. 

\begin{acknowledgments}
Part of this work was conducted at Advanced Characterization Nanotechnology Platform of the University of Tokyo, supported by ”Nanotechnology Platform” of the Ministry of Education, Culture, Sports, Science and Technology (MEXT), Japan. Sample characterization was performed by using a SQUID magnetometer at the CROSS user laboratory.  ARPES experiments were performed at UVSOR (Proposal No. 29-546). 
Preliminary experiment was done at KEK-PF under the approval of the Program Advisory Committee (Proposal Nos. 2015S2-003 and 2016G096).
This work was supported by KAKENHI Grant Nos. 15H02109 and 19K03741 from JSPS and by “Program for Promoting Researches on the Supercomputer Fugaku” (Basic Science for Emergence and Functionality in Quantum Matter, JPMXP1020200104) from MEXT.

\end{acknowledgments}


%


\begin{thebibliography}{23}%
	\makeatletter
	\providecommand \@ifxundefined [1]{%
		\@ifx{#1\undefined}
	}%
	\providecommand \@ifnum [1]{%
		\ifnum #1\expandafter \@firstoftwo
		\else \expandafter \@secondoftwo
		\fi
	}%
	\providecommand \@ifx [1]{%
		\ifx #1\expandafter \@firstoftwo
		\else \expandafter \@secondoftwo
		\fi
	}%
	\providecommand \natexlab [1]{#1}%
	\providecommand \enquote  [1]{``#1''}%
	\providecommand \bibnamefont  [1]{#1}%
	\providecommand \bibfnamefont [1]{#1}%
	\providecommand \citenamefont [1]{#1}%
	\providecommand \href@noop [0]{\@secondoftwo}%
	\providecommand \href [0]{\begingroup \@sanitize@url \@href}%
	\providecommand \@href[1]{\@@startlink{#1}\@@href}%
	\providecommand \@@href[1]{\endgroup#1\@@endlink}%
	\providecommand \@sanitize@url [0]{\catcode `\\12\catcode `\$12\catcode
		`\&12\catcode `\#12\catcode `\^12\catcode `\_12\catcode `\%12\relax}%
	\providecommand \@@startlink[1]{}%
	\providecommand \@@endlink[0]{}%
	\providecommand \url  [0]{\begingroup\@sanitize@url \@url }%
	\providecommand \@url [1]{\endgroup\@href {#1}{\urlprefix }}%
	\providecommand \urlprefix  [0]{URL }%
	\providecommand \Eprint [0]{\href }%
	\providecommand \doibase [0]{http://dx.doi.org/}%
	\providecommand \selectlanguage [0]{\@gobble}%
	\providecommand \bibinfo  [0]{\@secondoftwo}%
	\providecommand \bibfield  [0]{\@secondoftwo}%
	\providecommand \translation [1]{[#1]}%
	\providecommand \BibitemOpen [0]{}%
	\providecommand \bibitemStop [0]{}%
	\providecommand \bibitemNoStop [0]{.\EOS\space}%
	\providecommand \EOS [0]{\spacefactor3000\relax}%
	\providecommand \BibitemShut  [1]{\csname bibitem#1\endcsname}%
	\let\auto@bib@innerbib\@empty
	\bibitem [{\citenamefont {Kamihara}\ \emph {et~al.}(2008)\citenamefont
		{Kamihara}, \citenamefont {Watanabe}, \citenamefont {Hirano},\ and\
		\citenamefont {Hosono}}]{Kamihara2008}%
	\BibitemOpen
	\bibfield  {author} {\bibinfo {author} {\bibfnamefont {Y.}~\bibnamefont
			{Kamihara}}, \bibinfo {author} {\bibfnamefont {T.}~\bibnamefont {Watanabe}},
		\bibinfo {author} {\bibfnamefont {M.}~\bibnamefont {Hirano}}, \ and\ \bibinfo
		{author} {\bibfnamefont {H.}~\bibnamefont {Hosono}},\ }\href {\doibase
		10.1021/ja800073m} {\bibfield  {journal} {\bibinfo  {journal} {J. Am. Chem.
				Soc.}\ }\textbf {\bibinfo {volume} {130}},\ \bibinfo {pages} {3296} (\bibinfo
		{year} {2008})}\BibitemShut {NoStop}%
	\bibitem [{\citenamefont {Iimura}\ \emph {et~al.}(2012)\citenamefont {Iimura},
		\citenamefont {Matuishi}, \citenamefont {Sato}, \citenamefont {Hanna},
		\citenamefont {Muraba}, \citenamefont {Kim}, \citenamefont {Kim},
		\citenamefont {Takata},\ and\ \citenamefont {Hosono}}]{Iimura2012}%
	\BibitemOpen
	\bibfield  {author} {\bibinfo {author} {\bibfnamefont {S.}~\bibnamefont
			{Iimura}}, \bibinfo {author} {\bibfnamefont {S.}~\bibnamefont {Matuishi}},
		\bibinfo {author} {\bibfnamefont {H.}~\bibnamefont {Sato}}, \bibinfo {author}
		{\bibfnamefont {T.}~\bibnamefont {Hanna}}, \bibinfo {author} {\bibfnamefont
			{Y.}~\bibnamefont {Muraba}}, \bibinfo {author} {\bibfnamefont {S.~W.}\
			\bibnamefont {Kim}}, \bibinfo {author} {\bibfnamefont {J.~E.}\ \bibnamefont
			{Kim}}, \bibinfo {author} {\bibfnamefont {M.}~\bibnamefont {Takata}}, \ and\
		\bibinfo {author} {\bibfnamefont {H.}~\bibnamefont {Hosono}},\ }\href
	{\doibase 10.1038/ncomms1913} {\bibfield  {journal} {\bibinfo  {journal}
			{Nat. Commun.}\ }\textbf {\bibinfo {volume} {3}},\ \bibinfo {pages} {943}
		(\bibinfo {year} {2012})}\BibitemShut {NoStop}%
	\bibitem [{\citenamefont {Wang}\ \emph {et~al.}(2009)\citenamefont {Wang},
		\citenamefont {Li}, \citenamefont {Zhu}, \citenamefont {Jiang}, \citenamefont
		{Lin}, \citenamefont {Luo}, \citenamefont {Chi}, \citenamefont {Li},
		\citenamefont {Ren}, \citenamefont {He}, \citenamefont {Chen}, \citenamefont
		{Wang}, \citenamefont {Tao}, \citenamefont {Cao},\ and\ \citenamefont
		{Xu}}]{Wang2009}%
	\BibitemOpen
	\bibfield  {author} {\bibinfo {author} {\bibfnamefont {C.}~\bibnamefont
			{Wang}}, \bibinfo {author} {\bibfnamefont {Y.~K.}\ \bibnamefont {Li}},
		\bibinfo {author} {\bibfnamefont {Z.~W.}\ \bibnamefont {Zhu}}, \bibinfo
		{author} {\bibfnamefont {S.}~\bibnamefont {Jiang}}, \bibinfo {author}
		{\bibfnamefont {X.}~\bibnamefont {Lin}}, \bibinfo {author} {\bibfnamefont
			{Y.~K.}\ \bibnamefont {Luo}}, \bibinfo {author} {\bibfnamefont
			{S.}~\bibnamefont {Chi}}, \bibinfo {author} {\bibfnamefont {L.~J.}\
			\bibnamefont {Li}}, \bibinfo {author} {\bibfnamefont {Z.}~\bibnamefont
			{Ren}}, \bibinfo {author} {\bibfnamefont {M.}~\bibnamefont {He}}, \bibinfo
		{author} {\bibfnamefont {H.}~\bibnamefont {Chen}}, \bibinfo {author}
		{\bibfnamefont {Y.~T.}\ \bibnamefont {Wang}}, \bibinfo {author}
		{\bibfnamefont {Q.}~\bibnamefont {Tao}}, \bibinfo {author} {\bibfnamefont
			{G.~H.}\ \bibnamefont {Cao}}, \ and\ \bibinfo {author} {\bibfnamefont
			{Z.~A.}\ \bibnamefont {Xu}},\ }\href {\doibase 10.1103/PhysRevB.79.054521}
	{\bibfield  {journal} {\bibinfo  {journal} {Phys. Rev. B}\ }\textbf {\bibinfo
			{volume} {79}},\ \bibinfo {pages} {054521} (\bibinfo {year}
		{2009})}\BibitemShut {NoStop}%
	\bibitem [{\citenamefont {Lee}\ \emph {et~al.}(2008)\citenamefont {Lee},
		\citenamefont {Iyo}, \citenamefont {Eisaki}, \citenamefont {Kito},
		\citenamefont {{Teresa Fernandez-Diaz}}, \citenamefont {Ito}, \citenamefont
		{Kihou}, \citenamefont {Matsuhata}, \citenamefont {Braden},\ and\
		\citenamefont {Yamada}}]{Lee2008}%
	\BibitemOpen
	\bibfield  {author} {\bibinfo {author} {\bibfnamefont {C.-H.}\ \bibnamefont
			{Lee}}, \bibinfo {author} {\bibfnamefont {A.}~\bibnamefont {Iyo}}, \bibinfo
		{author} {\bibfnamefont {H.}~\bibnamefont {Eisaki}}, \bibinfo {author}
		{\bibfnamefont {H.}~\bibnamefont {Kito}}, \bibinfo {author} {\bibfnamefont
			{M.}~\bibnamefont {{Teresa Fernandez-Diaz}}}, \bibinfo {author}
		{\bibfnamefont {T.}~\bibnamefont {Ito}}, \bibinfo {author} {\bibfnamefont
			{K.}~\bibnamefont {Kihou}}, \bibinfo {author} {\bibfnamefont
			{H.}~\bibnamefont {Matsuhata}}, \bibinfo {author} {\bibfnamefont
			{M.}~\bibnamefont {Braden}}, \ and\ \bibinfo {author} {\bibfnamefont
			{K.}~\bibnamefont {Yamada}},\ }\href {\doibase 10.1143/JPSJ.77.083704}
	{\bibfield  {journal} {\bibinfo  {journal} {J. Phys. Soc. Jpn.}\ }\textbf
		{\bibinfo {volume} {77}},\ \bibinfo {pages} {083704} (\bibinfo {year}
		{2008})}\BibitemShut {NoStop}%
	\bibitem [{\citenamefont {Lu}\ \emph {et~al.}(2008)\citenamefont {Lu},
		\citenamefont {Yi}, \citenamefont {Mo}, \citenamefont {Erickson},
		\citenamefont {Analytis}, \citenamefont {Chu}, \citenamefont {Singh},
		\citenamefont {Hussain}, \citenamefont {Geballe}, \citenamefont {Fisher},\
		and\ \citenamefont {Shen}}]{Lu2008}%
	\BibitemOpen
	\bibfield  {author} {\bibinfo {author} {\bibfnamefont {D.~H.}\ \bibnamefont
			{Lu}}, \bibinfo {author} {\bibfnamefont {M.}~\bibnamefont {Yi}}, \bibinfo
		{author} {\bibfnamefont {S.-K.}\ \bibnamefont {Mo}}, \bibinfo {author}
		{\bibfnamefont {A.~S.}\ \bibnamefont {Erickson}}, \bibinfo {author}
		{\bibfnamefont {J.}~\bibnamefont {Analytis}}, \bibinfo {author}
		{\bibfnamefont {J.-H.}\ \bibnamefont {Chu}}, \bibinfo {author} {\bibfnamefont
			{D.~J.}\ \bibnamefont {Singh}}, \bibinfo {author} {\bibfnamefont
			{Z.}~\bibnamefont {Hussain}}, \bibinfo {author} {\bibfnamefont {T.~H.}\
			\bibnamefont {Geballe}}, \bibinfo {author} {\bibfnamefont {I.~R.}\
			\bibnamefont {Fisher}}, \ and\ \bibinfo {author} {\bibfnamefont {Z.-X.}\
			\bibnamefont {Shen}},\ }\href {\doibase 10.1038/nature07263} {\bibfield
		{journal} {\bibinfo  {journal} {Nature}\ }\textbf {\bibinfo {volume} {455}},\
		\bibinfo {pages} {81} (\bibinfo {year} {2008})}\BibitemShut {NoStop}%
	\bibitem [{\citenamefont {Nishi}\ \emph {et~al.}(2011)\citenamefont {Nishi},
		\citenamefont {Ishikado}, \citenamefont {Ideta}, \citenamefont {Malaeb},
		\citenamefont {Yoshida}, \citenamefont {Fujimori}, \citenamefont {Kotani},
		\citenamefont {Kubota}, \citenamefont {Ono}, \citenamefont {Yi},
		\citenamefont {Lu}, \citenamefont {Moore}, \citenamefont {Shen},
		\citenamefont {Iyo}, \citenamefont {Kihou}, \citenamefont {Kito},
		\citenamefont {Eisaki}, \citenamefont {Shamoto},\ and\ \citenamefont
		{Arita}}]{Nishi2011}%
	\BibitemOpen
	\bibfield  {author} {\bibinfo {author} {\bibfnamefont {I.}~\bibnamefont
			{Nishi}}, \bibinfo {author} {\bibfnamefont {M.}~\bibnamefont {Ishikado}},
		\bibinfo {author} {\bibfnamefont {S.}~\bibnamefont {Ideta}}, \bibinfo
		{author} {\bibfnamefont {W.}~\bibnamefont {Malaeb}}, \bibinfo {author}
		{\bibfnamefont {T.}~\bibnamefont {Yoshida}}, \bibinfo {author} {\bibfnamefont
			{A.}~\bibnamefont {Fujimori}}, \bibinfo {author} {\bibfnamefont
			{Y.}~\bibnamefont {Kotani}}, \bibinfo {author} {\bibfnamefont
			{M.}~\bibnamefont {Kubota}}, \bibinfo {author} {\bibfnamefont
			{K.}~\bibnamefont {Ono}}, \bibinfo {author} {\bibfnamefont {M.}~\bibnamefont
			{Yi}}, \bibinfo {author} {\bibfnamefont {D.~H.}\ \bibnamefont {Lu}}, \bibinfo
		{author} {\bibfnamefont {R.}~\bibnamefont {Moore}}, \bibinfo {author}
		{\bibfnamefont {Z.-X.}\ \bibnamefont {Shen}}, \bibinfo {author}
		{\bibfnamefont {A.}~\bibnamefont {Iyo}}, \bibinfo {author} {\bibfnamefont
			{K.}~\bibnamefont {Kihou}}, \bibinfo {author} {\bibfnamefont
			{H.}~\bibnamefont {Kito}}, \bibinfo {author} {\bibfnamefont {H.}~\bibnamefont
			{Eisaki}}, \bibinfo {author} {\bibfnamefont {S.}~\bibnamefont {Shamoto}}, \
		and\ \bibinfo {author} {\bibfnamefont {R.}~\bibnamefont {Arita}},\ }\href
	{\doibase 10.1103/PhysRevB.84.014504} {\bibfield  {journal} {\bibinfo
			{journal} {Phys. Rev. B}\ }\textbf {\bibinfo {volume} {84}},\ \bibinfo
		{pages} {014504} (\bibinfo {year} {2011})}\BibitemShut {NoStop}%
	\bibitem [{\citenamefont {Zhang}\ \emph {et~al.}(2016)\citenamefont {Zhang},
		\citenamefont {Ma}, \citenamefont {Qian}, \citenamefont {Shi}, \citenamefont
		{Fedorov}, \citenamefont {Denlinger}, \citenamefont {Wu}, \citenamefont {Hu},
		\citenamefont {Richard},\ and\ \citenamefont {Ding}}]{Zhang2016}%
	\BibitemOpen
	\bibfield  {author} {\bibinfo {author} {\bibfnamefont {P.}~\bibnamefont
			{Zhang}}, \bibinfo {author} {\bibfnamefont {J.}~\bibnamefont {Ma}}, \bibinfo
		{author} {\bibfnamefont {T.}~\bibnamefont {Qian}}, \bibinfo {author}
		{\bibfnamefont {Y.~G.}\ \bibnamefont {Shi}}, \bibinfo {author} {\bibfnamefont
			{A.~V.}\ \bibnamefont {Fedorov}}, \bibinfo {author} {\bibfnamefont {J.~D.}\
			\bibnamefont {Denlinger}}, \bibinfo {author} {\bibfnamefont {X.~X.}\
			\bibnamefont {Wu}}, \bibinfo {author} {\bibfnamefont {J.~P.}\ \bibnamefont
			{Hu}}, \bibinfo {author} {\bibfnamefont {P.}~\bibnamefont {Richard}}, \ and\
		\bibinfo {author} {\bibfnamefont {H.}~\bibnamefont {Ding}},\ }\href {\doibase
		10.1103/PhysRevB.94.104517} {\bibfield  {journal} {\bibinfo  {journal} {Phys.
				Rev. B}\ }\textbf {\bibinfo {volume} {94}},\ \bibinfo {pages} {104517}
		(\bibinfo {year} {2016})}\BibitemShut {NoStop}%
	\bibitem [{\citenamefont {Liu}\ \emph {et~al.}(2010)\citenamefont {Liu},
		\citenamefont {Lee}, \citenamefont {Palczewski}, \citenamefont {Yan},
		\citenamefont {Kondo}, \citenamefont {Harmon}, \citenamefont {McCallum},
		\citenamefont {Lograsso},\ and\ \citenamefont {Kaminski}}]{Liu2010}%
	\BibitemOpen
	\bibfield  {author} {\bibinfo {author} {\bibfnamefont {C.}~\bibnamefont
			{Liu}}, \bibinfo {author} {\bibfnamefont {Y.}~\bibnamefont {Lee}}, \bibinfo
		{author} {\bibfnamefont {A.~D.}\ \bibnamefont {Palczewski}}, \bibinfo
		{author} {\bibfnamefont {J.-Q.}\ \bibnamefont {Yan}}, \bibinfo {author}
		{\bibfnamefont {T.}~\bibnamefont {Kondo}}, \bibinfo {author} {\bibfnamefont
			{B.~N.}\ \bibnamefont {Harmon}}, \bibinfo {author} {\bibfnamefont {R.~W.}\
			\bibnamefont {McCallum}}, \bibinfo {author} {\bibfnamefont {T.~A.}\
			\bibnamefont {Lograsso}}, \ and\ \bibinfo {author} {\bibfnamefont
			{A.}~\bibnamefont {Kaminski}},\ }\href {\doibase 10.1103/PhysRevB.82.075135}
	{\bibfield  {journal} {\bibinfo  {journal} {Phys. Rev. B}\ }\textbf {\bibinfo
			{volume} {82}},\ \bibinfo {pages} {075135} (\bibinfo {year}
		{2010})}\BibitemShut {NoStop}%
	\bibitem [{\citenamefont {Charnukha}\ \emph {et~al.}(2015)\citenamefont
		{Charnukha}, \citenamefont {Thirupathaiah}, \citenamefont {Zabolotnyy},
		\citenamefont {B{\"{u}}chner}, \citenamefont {Zhigadlo}, \citenamefont
		{Batlogg}, \citenamefont {Yaresko},\ and\ \citenamefont
		{Borisenko}}]{Charnukha2015}%
	\BibitemOpen
	\bibfield  {author} {\bibinfo {author} {\bibfnamefont {A.}~\bibnamefont
			{Charnukha}}, \bibinfo {author} {\bibfnamefont {S.}~\bibnamefont
			{Thirupathaiah}}, \bibinfo {author} {\bibfnamefont {V.~B.}\ \bibnamefont
			{Zabolotnyy}}, \bibinfo {author} {\bibfnamefont {B.}~\bibnamefont
			{B{\"{u}}chner}}, \bibinfo {author} {\bibfnamefont {N.~D.}\ \bibnamefont
			{Zhigadlo}}, \bibinfo {author} {\bibfnamefont {B.}~\bibnamefont {Batlogg}},
		\bibinfo {author} {\bibfnamefont {A.~N.}\ \bibnamefont {Yaresko}}, \ and\
		\bibinfo {author} {\bibfnamefont {S.~V.}\ \bibnamefont {Borisenko}},\ }\href
	{\doibase 10.1038/srep10392} {\bibfield  {journal} {\bibinfo  {journal} {Sci.
				Rep.}\ }\textbf {\bibinfo {volume} {5}},\ \bibinfo {pages} {10392} (\bibinfo
		{year} {2015})}\BibitemShut {NoStop}%
	\bibitem [{\citenamefont {Charnukha}\ \emph {et~al.}(2016)\citenamefont
		{Charnukha}, \citenamefont {Evtushinsky}, \citenamefont {Matt}, \citenamefont
		{Xu}, \citenamefont {Shi}, \citenamefont {B{\"{u}}chner}, \citenamefont
		{Zhigadlo}, \citenamefont {Batlogg},\ and\ \citenamefont
		{Borisenko}}]{Charnukha2016}%
	\BibitemOpen
	\bibfield  {author} {\bibinfo {author} {\bibfnamefont {A.}~\bibnamefont
			{Charnukha}}, \bibinfo {author} {\bibfnamefont {D.~V.}\ \bibnamefont
			{Evtushinsky}}, \bibinfo {author} {\bibfnamefont {C.~E.}\ \bibnamefont
			{Matt}}, \bibinfo {author} {\bibfnamefont {N.}~\bibnamefont {Xu}}, \bibinfo
		{author} {\bibfnamefont {M.}~\bibnamefont {Shi}}, \bibinfo {author}
		{\bibfnamefont {B.}~\bibnamefont {B{\"{u}}chner}}, \bibinfo {author}
		{\bibfnamefont {N.~D.}\ \bibnamefont {Zhigadlo}}, \bibinfo {author}
		{\bibfnamefont {B.}~\bibnamefont {Batlogg}}, \ and\ \bibinfo {author}
		{\bibfnamefont {S.~V.}\ \bibnamefont {Borisenko}},\ }\href {\doibase
		10.1038/srep18273} {\bibfield  {journal} {\bibinfo  {journal} {Sci. Rep.}\
		}\textbf {\bibinfo {volume} {5}},\ \bibinfo {pages} {18273} (\bibinfo {year}
		{2016})}\BibitemShut {NoStop}%
	\bibitem [{\citenamefont {Kondo}\ \emph {et~al.}(2008)\citenamefont {Kondo},
		\citenamefont {Santander-Syro}, \citenamefont {Copie}, \citenamefont {Liu},
		\citenamefont {Tillman}, \citenamefont {Mun}, \citenamefont {Schmalian},
		\citenamefont {Bud'ko}, \citenamefont {Tanatar}, \citenamefont {Canfield},\
		and\ \citenamefont {Kaminski}}]{Kondo2008}%
	\BibitemOpen
	\bibfield  {author} {\bibinfo {author} {\bibfnamefont {T.}~\bibnamefont
			{Kondo}}, \bibinfo {author} {\bibfnamefont {A.~F.}\ \bibnamefont
			{Santander-Syro}}, \bibinfo {author} {\bibfnamefont {O.}~\bibnamefont
			{Copie}}, \bibinfo {author} {\bibfnamefont {C.}~\bibnamefont {Liu}}, \bibinfo
		{author} {\bibfnamefont {M.~E.}\ \bibnamefont {Tillman}}, \bibinfo {author}
		{\bibfnamefont {E.~D.}\ \bibnamefont {Mun}}, \bibinfo {author} {\bibfnamefont
			{J.}~\bibnamefont {Schmalian}}, \bibinfo {author} {\bibfnamefont {S.~L.}\
			\bibnamefont {Bud'ko}}, \bibinfo {author} {\bibfnamefont {M.~A.}\
			\bibnamefont {Tanatar}}, \bibinfo {author} {\bibfnamefont {P.~C.}\
			\bibnamefont {Canfield}}, \ and\ \bibinfo {author} {\bibfnamefont
			{A.}~\bibnamefont {Kaminski}},\ }\href {\doibase
		10.1103/PhysRevLett.101.147003} {\bibfield  {journal} {\bibinfo  {journal}
			{Phys. Rev. Lett.}\ }\textbf {\bibinfo {volume} {101}},\ \bibinfo {pages}
		{147003} (\bibinfo {year} {2008})}\BibitemShut {NoStop}%
	\bibitem [{\citenamefont {Sato}\ \emph
		{et~al.}(2008{\natexlab{a}})\citenamefont {Sato}, \citenamefont {Souma},
		\citenamefont {Nakayama}, \citenamefont {Terashima}, \citenamefont
		{Sugawara}, \citenamefont {Takahashi}, \citenamefont {Kamihara},
		\citenamefont {Hirano},\ and\ \citenamefont {Hosono}}]{Sato2008}%
	\BibitemOpen
	\bibfield  {author} {\bibinfo {author} {\bibfnamefont {T.}~\bibnamefont
			{Sato}}, \bibinfo {author} {\bibfnamefont {S.}~\bibnamefont {Souma}},
		\bibinfo {author} {\bibfnamefont {K.}~\bibnamefont {Nakayama}}, \bibinfo
		{author} {\bibfnamefont {K.}~\bibnamefont {Terashima}}, \bibinfo {author}
		{\bibfnamefont {K.}~\bibnamefont {Sugawara}}, \bibinfo {author}
		{\bibfnamefont {T.}~\bibnamefont {Takahashi}}, \bibinfo {author}
		{\bibfnamefont {Y.}~\bibnamefont {Kamihara}}, \bibinfo {author}
		{\bibfnamefont {M.}~\bibnamefont {Hirano}}, \ and\ \bibinfo {author}
		{\bibfnamefont {H.}~\bibnamefont {Hosono}},\ }\href {\doibase
		10.1143/JPSJ.77.063708} {\bibfield  {journal} {\bibinfo  {journal} {J. Phys.
				Soc. Jpn.}\ }\textbf {\bibinfo {volume} {77}},\ \bibinfo {pages} {063708}
		(\bibinfo {year} {2008}{\natexlab{a}})}\BibitemShut {NoStop}%
	\bibitem [{\citenamefont {Sato}\ \emph
		{et~al.}(2008{\natexlab{b}})\citenamefont {Sato}, \citenamefont {Nakayama},
		\citenamefont {Sekiba}, \citenamefont {Arakane}, \citenamefont {Terashima},
		\citenamefont {Souma}, \citenamefont {Takahashi}, \citenamefont {Kamihara},
		\citenamefont {Hirano},\ and\ \citenamefont {Hosono}}]{Sato2008a}%
	\BibitemOpen
	\bibfield  {author} {\bibinfo {author} {\bibfnamefont {T.}~\bibnamefont
			{Sato}}, \bibinfo {author} {\bibfnamefont {K.}~\bibnamefont {Nakayama}},
		\bibinfo {author} {\bibfnamefont {Y.}~\bibnamefont {Sekiba}}, \bibinfo
		{author} {\bibfnamefont {T.}~\bibnamefont {Arakane}}, \bibinfo {author}
		{\bibfnamefont {K.}~\bibnamefont {Terashima}}, \bibinfo {author}
		{\bibfnamefont {S.}~\bibnamefont {Souma}}, \bibinfo {author} {\bibfnamefont
			{T.}~\bibnamefont {Takahashi}}, \bibinfo {author} {\bibfnamefont
			{Y.}~\bibnamefont {Kamihara}}, \bibinfo {author} {\bibfnamefont
			{M.}~\bibnamefont {Hirano}}, \ and\ \bibinfo {author} {\bibfnamefont
			{H.}~\bibnamefont {Hosono}},\ }\href {\doibase 10.1143/JPSJS.77SC.65}
	{\bibfield  {journal} {\bibinfo  {journal} {J. Phys. Soc. Jpn.}\ }\textbf
		{\bibinfo {volume} {77}},\ \bibinfo {pages} {65} (\bibinfo {year}
		{2008}{\natexlab{b}})}\BibitemShut {NoStop}%
	\bibitem [{\citenamefont {Nakamura}\ \emph {et~al.}(2017)\citenamefont
		{Nakamura}, \citenamefont {Shimojima}, \citenamefont {Sonobe}, \citenamefont
		{Yoshida}, \citenamefont {Ishizaka}, \citenamefont {Malaeb}, \citenamefont
		{Shin}, \citenamefont {Iimura}, \citenamefont {Matsuishi},\ and\
		\citenamefont {Hosono}}]{Nakamura2017}%
	\BibitemOpen
	\bibfield  {author} {\bibinfo {author} {\bibfnamefont {A.}~\bibnamefont
			{Nakamura}}, \bibinfo {author} {\bibfnamefont {T.}~\bibnamefont {Shimojima}},
		\bibinfo {author} {\bibfnamefont {T.}~\bibnamefont {Sonobe}}, \bibinfo
		{author} {\bibfnamefont {S.}~\bibnamefont {Yoshida}}, \bibinfo {author}
		{\bibfnamefont {K.}~\bibnamefont {Ishizaka}}, \bibinfo {author}
		{\bibfnamefont {W.}~\bibnamefont {Malaeb}}, \bibinfo {author} {\bibfnamefont
			{S.}~\bibnamefont {Shin}}, \bibinfo {author} {\bibfnamefont {S.}~\bibnamefont
			{Iimura}}, \bibinfo {author} {\bibfnamefont {S.}~\bibnamefont {Matsuishi}}, \
		and\ \bibinfo {author} {\bibfnamefont {H.}~\bibnamefont {Hosono}},\ }\href
	{\doibase 10.1103/PhysRevB.95.064501} {\bibfield  {journal} {\bibinfo
			{journal} {Phys. Rev. B}\ }\textbf {\bibinfo {volume} {95}},\ \bibinfo
		{pages} {064501} (\bibinfo {year} {2017})}\BibitemShut {NoStop}%
	\bibitem [{\citenamefont {Ishikado}\ \emph {et~al.}(2009)\citenamefont
		{Ishikado}, \citenamefont {Shamoto}, \citenamefont {Kito}, \citenamefont
		{Iyo}, \citenamefont {Eisaki}, \citenamefont {Ito},\ and\ \citenamefont
		{Tomioka}}]{Ishikado2009}%
	\BibitemOpen
	\bibfield  {author} {\bibinfo {author} {\bibfnamefont {M.}~\bibnamefont
			{Ishikado}}, \bibinfo {author} {\bibfnamefont {S.}~\bibnamefont {Shamoto}},
		\bibinfo {author} {\bibfnamefont {H.}~\bibnamefont {Kito}}, \bibinfo {author}
		{\bibfnamefont {A.}~\bibnamefont {Iyo}}, \bibinfo {author} {\bibfnamefont
			{H.}~\bibnamefont {Eisaki}}, \bibinfo {author} {\bibfnamefont
			{T.}~\bibnamefont {Ito}}, \ and\ \bibinfo {author} {\bibfnamefont
			{Y.}~\bibnamefont {Tomioka}},\ }\href {\doibase 10.1016/j.physc.2009.05.094}
	{\bibfield  {journal} {\bibinfo  {journal} {Physica C}\ }\textbf {\bibinfo
			{volume} {469}},\ \bibinfo {pages} {901} (\bibinfo {year}
		{2009})}\BibitemShut {NoStop}%
	\bibitem [{\citenamefont {Ishikado}\ \emph {et~al.}(2010)\citenamefont
		{Ishikado}, \citenamefont {Shamoto}, \citenamefont {Kito}, \citenamefont
		{Iyo}, \citenamefont {Eisaki}, \citenamefont {Ito},\ and\ \citenamefont
		{Tomioka}}]{ishikado2010growth}%
	\BibitemOpen
	\bibfield  {author} {\bibinfo {author} {\bibfnamefont {M.}~\bibnamefont
			{Ishikado}}, \bibinfo {author} {\bibfnamefont {S.}~\bibnamefont {Shamoto}},
		\bibinfo {author} {\bibfnamefont {H.}~\bibnamefont {Kito}}, \bibinfo {author}
		{\bibfnamefont {A.}~\bibnamefont {Iyo}}, \bibinfo {author} {\bibfnamefont
			{H.}~\bibnamefont {Eisaki}}, \bibinfo {author} {\bibfnamefont
			{T.}~\bibnamefont {Ito}}, \ and\ \bibinfo {author} {\bibfnamefont
			{Y.}~\bibnamefont {Tomioka}},\ }\href
	{https://www.sciencedirect.com/science/article/pii/S092145340900882X}
	{\bibfield  {journal} {\bibinfo  {journal} {Physica C}\ }\textbf {\bibinfo
			{volume} {470}},\ \bibinfo {pages} {S322} (\bibinfo {year}
		{2010})}\BibitemShut {NoStop}%
	\bibitem [{\citenamefont {Okazaki}\ \emph {et~al.}(2009)\citenamefont
		{Okazaki}, \citenamefont {Konczykowski}, \citenamefont {van~der Beek},
		\citenamefont {Kato}, \citenamefont {Hashimoto}, \citenamefont {Shimozawa},
		\citenamefont {Shishido}, \citenamefont {Yamashita}, \citenamefont
		{Ishikado}, \citenamefont {Kito}, \citenamefont {Iyo}, \citenamefont
		{Eisaki}, \citenamefont {Shamoto}, \citenamefont {Shibauchi},\ and\
		\citenamefont {Matsuda}}]{okazaki2009lower}%
	\BibitemOpen
	\bibfield  {author} {\bibinfo {author} {\bibfnamefont {R.}~\bibnamefont
			{Okazaki}}, \bibinfo {author} {\bibfnamefont {M.}~\bibnamefont
			{Konczykowski}}, \bibinfo {author} {\bibfnamefont {C.~J.}\ \bibnamefont
			{van~der Beek}}, \bibinfo {author} {\bibfnamefont {T.}~\bibnamefont {Kato}},
		\bibinfo {author} {\bibfnamefont {K.}~\bibnamefont {Hashimoto}}, \bibinfo
		{author} {\bibfnamefont {M.}~\bibnamefont {Shimozawa}}, \bibinfo {author}
		{\bibfnamefont {H.}~\bibnamefont {Shishido}}, \bibinfo {author}
		{\bibfnamefont {M.}~\bibnamefont {Yamashita}}, \bibinfo {author}
		{\bibfnamefont {M.}~\bibnamefont {Ishikado}}, \bibinfo {author}
		{\bibfnamefont {H.}~\bibnamefont {Kito}}, \bibinfo {author} {\bibfnamefont
			{A.}~\bibnamefont {Iyo}}, \bibinfo {author} {\bibfnamefont {H.}~\bibnamefont
			{Eisaki}}, \bibinfo {author} {\bibfnamefont {S.}~\bibnamefont {Shamoto}},
		\bibinfo {author} {\bibfnamefont {T.}~\bibnamefont {Shibauchi}}, \ and\
		\bibinfo {author} {\bibfnamefont {Y.}~\bibnamefont {Matsuda}},\ }\href
	{\doibase 10.1103/PhysRevB.79.064520} {\bibfield  {journal} {\bibinfo
			{journal} {Phys. Rev. B}\ }\textbf {\bibinfo {volume} {79}},\ \bibinfo
		{pages} {064520} (\bibinfo {year} {2009})}\BibitemShut {NoStop}%
	\bibitem [{\citenamefont {van~der Beek}\ \emph {et~al.}(2010)\citenamefont
		{van~der Beek}, \citenamefont {Rizza}, \citenamefont {Konczykowski},
		\citenamefont {Fertey}, \citenamefont {Monnet}, \citenamefont {Klein},
		\citenamefont {Okazaki}, \citenamefont {Ishikado}, \citenamefont {Kito},
		\citenamefont {Iyo}, \citenamefont {Eisaki}, \citenamefont {Shamoto},
		\citenamefont {Tillman}, \citenamefont {Bud'ko}, \citenamefont {Canfield},
		\citenamefont {Shibauchi},\ and\ \citenamefont {Matsuda}}]{van2010flux}%
	\BibitemOpen
	\bibfield  {author} {\bibinfo {author} {\bibfnamefont {C.~J.}\ \bibnamefont
			{van~der Beek}}, \bibinfo {author} {\bibfnamefont {G.}~\bibnamefont {Rizza}},
		\bibinfo {author} {\bibfnamefont {M.}~\bibnamefont {Konczykowski}}, \bibinfo
		{author} {\bibfnamefont {P.}~\bibnamefont {Fertey}}, \bibinfo {author}
		{\bibfnamefont {I.}~\bibnamefont {Monnet}}, \bibinfo {author} {\bibfnamefont
			{T.}~\bibnamefont {Klein}}, \bibinfo {author} {\bibfnamefont
			{R.}~\bibnamefont {Okazaki}}, \bibinfo {author} {\bibfnamefont
			{M.}~\bibnamefont {Ishikado}}, \bibinfo {author} {\bibfnamefont
			{H.}~\bibnamefont {Kito}}, \bibinfo {author} {\bibfnamefont {A.}~\bibnamefont
			{Iyo}}, \bibinfo {author} {\bibfnamefont {H.}~\bibnamefont {Eisaki}},
		\bibinfo {author} {\bibfnamefont {S.}~\bibnamefont {Shamoto}}, \bibinfo
		{author} {\bibfnamefont {M.~E.}\ \bibnamefont {Tillman}}, \bibinfo {author}
		{\bibfnamefont {S.~L.}\ \bibnamefont {Bud'ko}}, \bibinfo {author}
		{\bibfnamefont {P.~C.}\ \bibnamefont {Canfield}}, \bibinfo {author}
		{\bibfnamefont {T.}~\bibnamefont {Shibauchi}}, \ and\ \bibinfo {author}
		{\bibfnamefont {Y.}~\bibnamefont {Matsuda}},\ }\href {\doibase
		10.1103/PhysRevB.81.174517} {\bibfield  {journal} {\bibinfo  {journal} {Phys.
				Rev. B}\ }\textbf {\bibinfo {volume} {81}},\ \bibinfo {pages} {174517}
		(\bibinfo {year} {2010})}\BibitemShut {NoStop}%
	\bibitem [{\citenamefont {Ding}\ \emph {et~al.}(2008)\citenamefont {Ding},
		\citenamefont {Richard}, \citenamefont {Nakayama}, \citenamefont {Sugawara},
		\citenamefont {Arakane}, \citenamefont {Sekiba}, \citenamefont {Takayama},
		\citenamefont {Souma}, \citenamefont {Sato}, \citenamefont {Takahashi},
		\citenamefont {Wang}, \citenamefont {Dai}, \citenamefont {Fang},
		\citenamefont {Chen}, \citenamefont {Luo},\ and\ \citenamefont
		{Wang}}]{Ding2008}%
	\BibitemOpen
	\bibfield  {author} {\bibinfo {author} {\bibfnamefont {H.}~\bibnamefont
			{Ding}}, \bibinfo {author} {\bibfnamefont {P.}~\bibnamefont {Richard}},
		\bibinfo {author} {\bibfnamefont {K.}~\bibnamefont {Nakayama}}, \bibinfo
		{author} {\bibfnamefont {K.}~\bibnamefont {Sugawara}}, \bibinfo {author}
		{\bibfnamefont {T.}~\bibnamefont {Arakane}}, \bibinfo {author} {\bibfnamefont
			{Y.}~\bibnamefont {Sekiba}}, \bibinfo {author} {\bibfnamefont
			{A.}~\bibnamefont {Takayama}}, \bibinfo {author} {\bibfnamefont
			{S.}~\bibnamefont {Souma}}, \bibinfo {author} {\bibfnamefont
			{T.}~\bibnamefont {Sato}}, \bibinfo {author} {\bibfnamefont {T.}~\bibnamefont
			{Takahashi}}, \bibinfo {author} {\bibfnamefont {Z.}~\bibnamefont {Wang}},
		\bibinfo {author} {\bibfnamefont {X.}~\bibnamefont {Dai}}, \bibinfo {author}
		{\bibfnamefont {Z.}~\bibnamefont {Fang}}, \bibinfo {author} {\bibfnamefont
			{G.~F.}\ \bibnamefont {Chen}}, \bibinfo {author} {\bibfnamefont {J.~L.}\
			\bibnamefont {Luo}}, \ and\ \bibinfo {author} {\bibfnamefont {N.~L.}\
			\bibnamefont {Wang}},\ }\href {\doibase 10.1209/0295-5075/83/47001}
	{\bibfield  {journal} {\bibinfo  {journal} {Europhys. Lett.}\ }\textbf
		{\bibinfo {volume} {83}},\ \bibinfo {pages} {47001} (\bibinfo {year}
		{2008})}\BibitemShut {NoStop}%
	\bibitem [{\citenamefont {Okazaki}\ \emph {et~al.}(2012)\citenamefont
		{Okazaki}, \citenamefont {Ota}, \citenamefont {Kotani}, \citenamefont
		{Malaeb}, \citenamefont {Ishida}, \citenamefont {Shimojima}, \citenamefont
		{Kiss}, \citenamefont {Watanabe}, \citenamefont {Chen}, \citenamefont
		{Kihou}, \citenamefont {Lee}, \citenamefont {Iyo}, \citenamefont {Eisaki},
		\citenamefont {Saito}, \citenamefont {Fukazawa}, \citenamefont {Kohori},
		\citenamefont {Hashimoto}, \citenamefont {Shibauchi}, \citenamefont
		{Matsuda},\ and\ \citenamefont {Shin}}]{okazaki2012octet}%
	\BibitemOpen
	\bibfield  {author} {\bibinfo {author} {\bibfnamefont {K.}~\bibnamefont
			{Okazaki}}, \bibinfo {author} {\bibfnamefont {Y.}~\bibnamefont {Ota}},
		\bibinfo {author} {\bibfnamefont {Y.}~\bibnamefont {Kotani}}, \bibinfo
		{author} {\bibfnamefont {W.}~\bibnamefont {Malaeb}}, \bibinfo {author}
		{\bibfnamefont {Y.}~\bibnamefont {Ishida}}, \bibinfo {author} {\bibfnamefont
			{T.}~\bibnamefont {Shimojima}}, \bibinfo {author} {\bibfnamefont
			{T.}~\bibnamefont {Kiss}}, \bibinfo {author} {\bibfnamefont {S.}~\bibnamefont
			{Watanabe}}, \bibinfo {author} {\bibfnamefont {C.-T.}\ \bibnamefont {Chen}},
		\bibinfo {author} {\bibfnamefont {K.}~\bibnamefont {Kihou}}, \bibinfo
		{author} {\bibfnamefont {C.-H.}\ \bibnamefont {Lee}}, \bibinfo {author}
		{\bibfnamefont {A.}~\bibnamefont {Iyo}}, \bibinfo {author} {\bibfnamefont
			{H.}~\bibnamefont {Eisaki}}, \bibinfo {author} {\bibfnamefont
			{T.}~\bibnamefont {Saito}}, \bibinfo {author} {\bibfnamefont
			{H.}~\bibnamefont {Fukazawa}}, \bibinfo {author} {\bibfnamefont
			{Y.}~\bibnamefont {Kohori}}, \bibinfo {author} {\bibfnamefont
			{K.}~\bibnamefont {Hashimoto}}, \bibinfo {author} {\bibfnamefont
			{T.}~\bibnamefont {Shibauchi}}, \bibinfo {author} {\bibfnamefont
			{Y.}~\bibnamefont {Matsuda}}, \ and\ \bibinfo {author} {\bibfnamefont
			{S.}~\bibnamefont {Shin}},\ }\href {\doibase 10.1126/science.1222793}
	{\bibfield  {journal} {\bibinfo  {journal} {Science}\ }\textbf {\bibinfo
			{volume} {337}},\ \bibinfo {pages} {1314} (\bibinfo {year}
		{2012})}\BibitemShut {NoStop}%
	\bibitem [{\citenamefont {Malaeb}\ \emph {et~al.}(2012)\citenamefont {Malaeb},
		\citenamefont {Shimojima}, \citenamefont {Ishida}, \citenamefont {Okazaki},
		\citenamefont {Ota}, \citenamefont {Ohgushi}, \citenamefont {Kihou},
		\citenamefont {Saito}, \citenamefont {Lee}, \citenamefont {Ishida},
		\citenamefont {Nakajima}, \citenamefont {Uchida}, \citenamefont {Fukazawa},
		\citenamefont {Kohori}, \citenamefont {Iyo}, \citenamefont {Eisaki},
		\citenamefont {Chen}, \citenamefont {Watanabe}, \citenamefont {Ikeda},\ and\
		\citenamefont {Shin}}]{Malaeb2012}%
	\BibitemOpen
	\bibfield  {author} {\bibinfo {author} {\bibfnamefont {W.}~\bibnamefont
			{Malaeb}}, \bibinfo {author} {\bibfnamefont {T.}~\bibnamefont {Shimojima}},
		\bibinfo {author} {\bibfnamefont {Y.}~\bibnamefont {Ishida}}, \bibinfo
		{author} {\bibfnamefont {K.}~\bibnamefont {Okazaki}}, \bibinfo {author}
		{\bibfnamefont {Y.}~\bibnamefont {Ota}}, \bibinfo {author} {\bibfnamefont
			{K.}~\bibnamefont {Ohgushi}}, \bibinfo {author} {\bibfnamefont
			{K.}~\bibnamefont {Kihou}}, \bibinfo {author} {\bibfnamefont
			{T.}~\bibnamefont {Saito}}, \bibinfo {author} {\bibfnamefont {C.~H.}\
			\bibnamefont {Lee}}, \bibinfo {author} {\bibfnamefont {S.}~\bibnamefont
			{Ishida}}, \bibinfo {author} {\bibfnamefont {M.}~\bibnamefont {Nakajima}},
		\bibinfo {author} {\bibfnamefont {S.}~\bibnamefont {Uchida}}, \bibinfo
		{author} {\bibfnamefont {H.}~\bibnamefont {Fukazawa}}, \bibinfo {author}
		{\bibfnamefont {Y.}~\bibnamefont {Kohori}}, \bibinfo {author} {\bibfnamefont
			{A.}~\bibnamefont {Iyo}}, \bibinfo {author} {\bibfnamefont {H.}~\bibnamefont
			{Eisaki}}, \bibinfo {author} {\bibfnamefont {C.-T.}\ \bibnamefont {Chen}},
		\bibinfo {author} {\bibfnamefont {S.}~\bibnamefont {Watanabe}}, \bibinfo
		{author} {\bibfnamefont {H.}~\bibnamefont {Ikeda}}, \ and\ \bibinfo {author}
		{\bibfnamefont {S.}~\bibnamefont {Shin}},\ }\href {\doibase
		10.1103/PhysRevB.86.165117} {\bibfield  {journal} {\bibinfo  {journal} {Phys.
				Rev. B}\ }\textbf {\bibinfo {volume} {86}},\ \bibinfo {pages} {165117}
		(\bibinfo {year} {2012})}\BibitemShut {NoStop}%
	\bibitem [{\citenamefont {Xu}\ \emph {et~al.}(2011)\citenamefont {Xu},
		\citenamefont {Richard}, \citenamefont {Nakayama}, \citenamefont {Kawahara},
		\citenamefont {Sekiba}, \citenamefont {Qian}, \citenamefont {Neupane},
		\citenamefont {Souma}, \citenamefont {Sato}, \citenamefont {Takahashi},
		\citenamefont {Luo}, \citenamefont {Wen}, \citenamefont {Chen}, \citenamefont
		{Wang}, \citenamefont {Wang}, \citenamefont {Fang}, \citenamefont {Dai},\
		and\ \citenamefont {Ding}}]{Xu2011}%
	\BibitemOpen
	\bibfield  {author} {\bibinfo {author} {\bibfnamefont {Y.-M.}\ \bibnamefont
			{Xu}}, \bibinfo {author} {\bibfnamefont {P.}~\bibnamefont {Richard}},
		\bibinfo {author} {\bibfnamefont {K.}~\bibnamefont {Nakayama}}, \bibinfo
		{author} {\bibfnamefont {T.}~\bibnamefont {Kawahara}}, \bibinfo {author}
		{\bibfnamefont {Y.}~\bibnamefont {Sekiba}}, \bibinfo {author} {\bibfnamefont
			{T.}~\bibnamefont {Qian}}, \bibinfo {author} {\bibfnamefont {M.}~\bibnamefont
			{Neupane}}, \bibinfo {author} {\bibfnamefont {S.}~\bibnamefont {Souma}},
		\bibinfo {author} {\bibfnamefont {T.}~\bibnamefont {Sato}}, \bibinfo {author}
		{\bibfnamefont {T.}~\bibnamefont {Takahashi}}, \bibinfo {author}
		{\bibfnamefont {H.-Q.}\ \bibnamefont {Luo}}, \bibinfo {author} {\bibfnamefont
			{H.-H.}\ \bibnamefont {Wen}}, \bibinfo {author} {\bibfnamefont {G.-F.}\
			\bibnamefont {Chen}}, \bibinfo {author} {\bibfnamefont {N.-L.}\ \bibnamefont
			{Wang}}, \bibinfo {author} {\bibfnamefont {Z.}~\bibnamefont {Wang}}, \bibinfo
		{author} {\bibfnamefont {Z.}~\bibnamefont {Fang}}, \bibinfo {author}
		{\bibfnamefont {X.}~\bibnamefont {Dai}}, \ and\ \bibinfo {author}
		{\bibfnamefont {H.}~\bibnamefont {Ding}},\ }\href {\doibase
		10.1038/ncomms1394} {\bibfield  {journal} {\bibinfo  {journal} {Nat.
				Commun.}\ }\textbf {\bibinfo {volume} {2}},\ \bibinfo {pages} {394} (\bibinfo
		{year} {2011})}\BibitemShut {NoStop}%
	\bibitem [{\citenamefont {Kwon}\ \emph {et~al.}(2012)\citenamefont {Kwon},
		\citenamefont {Hong}, \citenamefont {Jang}, \citenamefont {Oh}, \citenamefont
		{Song}, \citenamefont {Min}, \citenamefont {Iizuka}, \citenamefont {Kimura},
		\citenamefont {Balatsky},\ and\ \citenamefont {Bang}}]{kwon2012evidence}%
	\BibitemOpen
	\bibfield  {author} {\bibinfo {author} {\bibfnamefont {Y.~S.}\ \bibnamefont
			{Kwon}}, \bibinfo {author} {\bibfnamefont {J.~B.}\ \bibnamefont {Hong}},
		\bibinfo {author} {\bibfnamefont {Y.~R.}\ \bibnamefont {Jang}}, \bibinfo
		{author} {\bibfnamefont {H.~J.}\ \bibnamefont {Oh}}, \bibinfo {author}
		{\bibfnamefont {Y.~Y.}\ \bibnamefont {Song}}, \bibinfo {author}
		{\bibfnamefont {B.~H.}\ \bibnamefont {Min}}, \bibinfo {author} {\bibfnamefont
			{T.}~\bibnamefont {Iizuka}}, \bibinfo {author} {\bibfnamefont {S.-i.}\
			\bibnamefont {Kimura}}, \bibinfo {author} {\bibfnamefont {A.}~\bibnamefont
			{Balatsky}}, \ and\ \bibinfo {author} {\bibfnamefont {Y.}~\bibnamefont
			{Bang}},\ }\href
	{https://iopscience.iop.org/article/10.1088/1367-2630/14/6/063009/meta}
	{\bibfield  {journal} {\bibinfo  {journal} {New J. Phys.}\ }\textbf {\bibinfo
			{volume} {14}},\ \bibinfo {pages} {063009} (\bibinfo {year}
		{2012})}\BibitemShut {NoStop}%
\end{thebibliography}

\clearpage

\section*{Supplementary information}
We made analysis by taking spectra at 97/94K  (b, f, j), 84/80K (c, g, k), and 73/70K (d, h, l), in addition to 117/105K (a, e, i) as the normal-state EDC in Fig. S \ref{fig:S1}.
As one can see from Figs. S \ref{fig:S2} and S \ref{fig:S3}, normalized gap areas obtained using different normal-state EDCs behave nearly in the same way.
Thus, our results are independent of the normal-state EDC at different temperatures.
 
In order to strengthen our conclusion derived from the analysis of subtracting the featureless EDCs as the background, we have also tested another type of background.
Here, we consider Fermi-Dirac-function(FD)-divided EDCs.
As shown in Fig. S \ref{fig:S4} (a), we take a linear background whose slope is determined by fitting the FD-divided EDC in the high binding energy region ($E-E_\text{F}\leq-35$ meV). 
One can recognize a sharp edge after background subtraction indicated by a black triangle.
The size of the sharp edge is nearly equal to the gap size we estimated in the main text, and can be associated with the superconducting gap.
Figures S \ref{fig:S4} (c), (e), and (f) demonstrate the temperature dependences of the background-subtracted FD-divided EDCs on the Outer, Middle, and Inner hole pockets, respectively.
Black dotted lines indicate a base line in the high binding energy region.
At higher temperatures, the spectra are parallel to the black dotted lines in the entire energy range, indicating a closure of the pseudogap.
As shown in figs. S \ref{fig:S5} and S \ref{fig:S6}, one finds that the gap area roughly corresponds to those estimated by the different way described in the main text.


\setcounter{figure}{0}
\renewcommand{\figurename}{FIG. S}

\begin{figure*}
\includegraphics[clip,width=12cm]{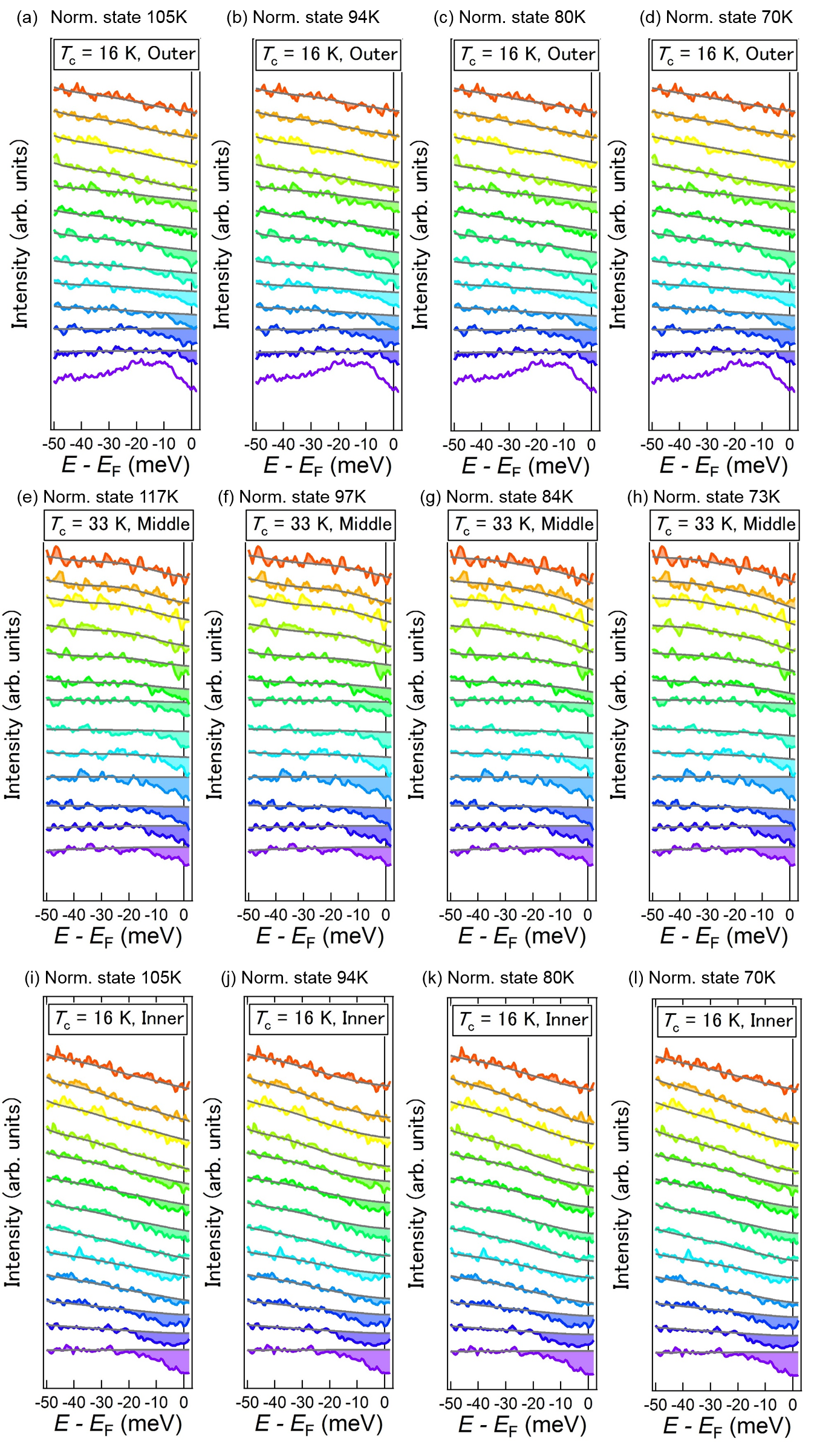}
\caption{
Analysis using different normal-state EDCs. We take a spectrum at 117/105K (a, e, i), 97/94K  (b, f, j), 84/80K (c, g, k), and 73/70K (d, h, l) as the normal-state EDC.
}\label{fig:S1}
\end{figure*}

\begin{figure*}
\includegraphics[clip,width=14cm]{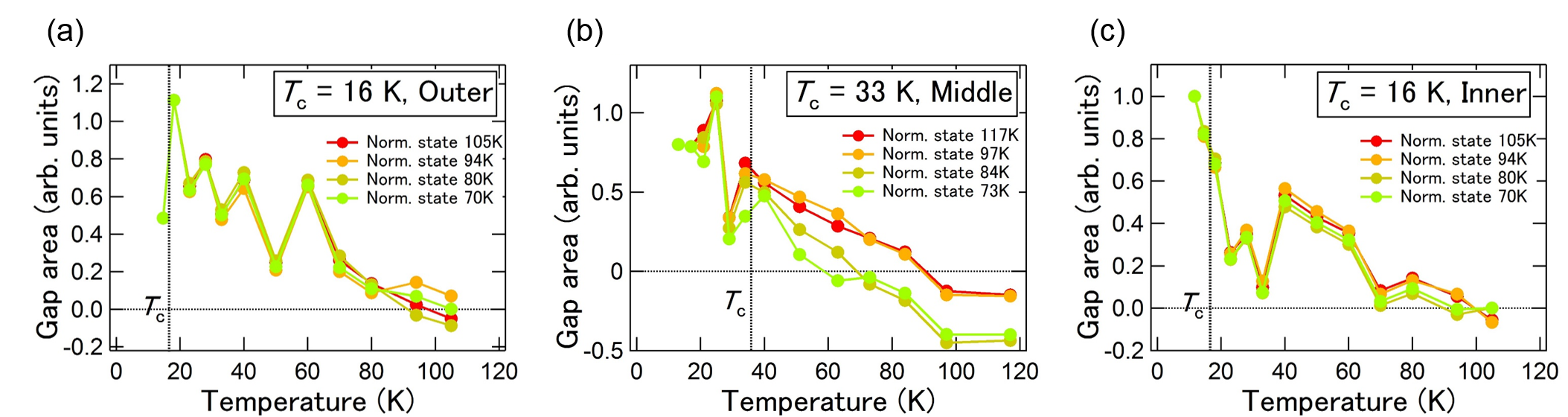}
\caption{
Gap area obtained from Fig. S \ref{fig:S1}..
}\label{fig:S2}
\end{figure*}

\begin{figure*}
\includegraphics[clip,width=14cm]{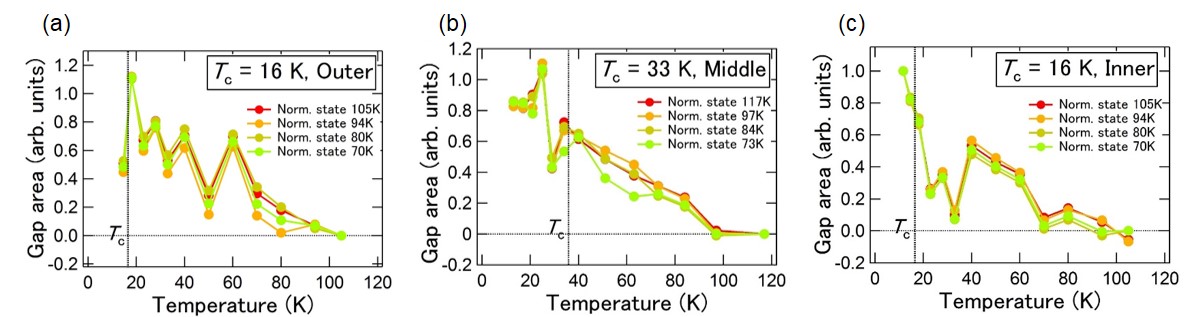}
\caption{
Normalized gap area obtained from Fig. S \ref{fig:S1}.
}\label{fig:S3}
\end{figure*}

\begin{figure*}
\includegraphics[clip,width=14cm]{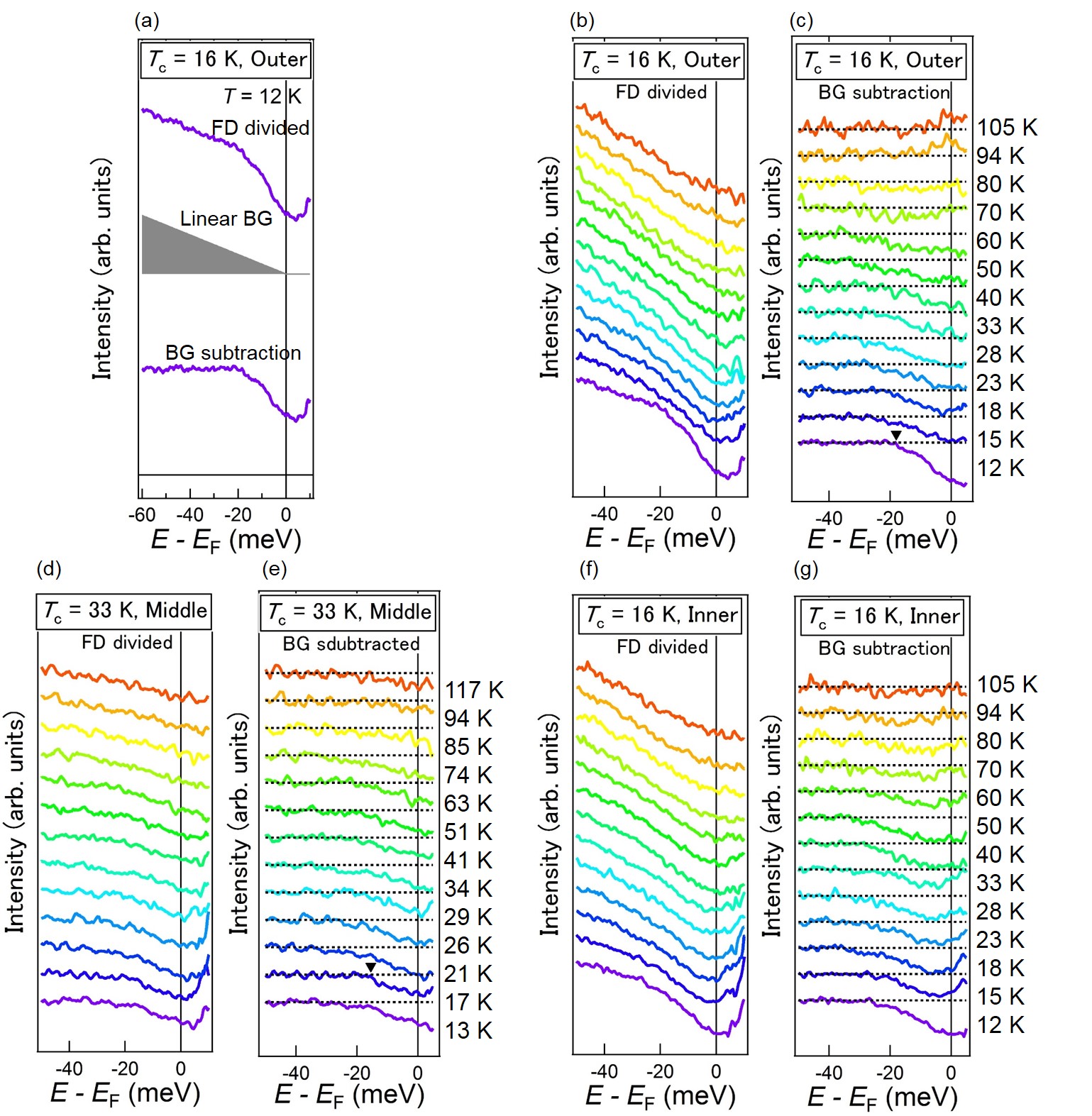}
\caption{
(a) Fermi-Dirac-function-divided EDCs before and after the linear background subtraction on the Outer hole pocket at $T = 12$ K .
Temperature dependence of the Fermi-Dirac-function-divided EDCs (b, d, f) and the background-subtracted EDCs (c, e, g) on the Outer (b, c), Middle (d, e), and Inner (f, g) hole pockets.
Black triangles indicate a sharp edge associated with the superconducting gap.
}\label{fig:S4}
\end{figure*}

\begin{figure*}
\includegraphics[clip,width=14cm]{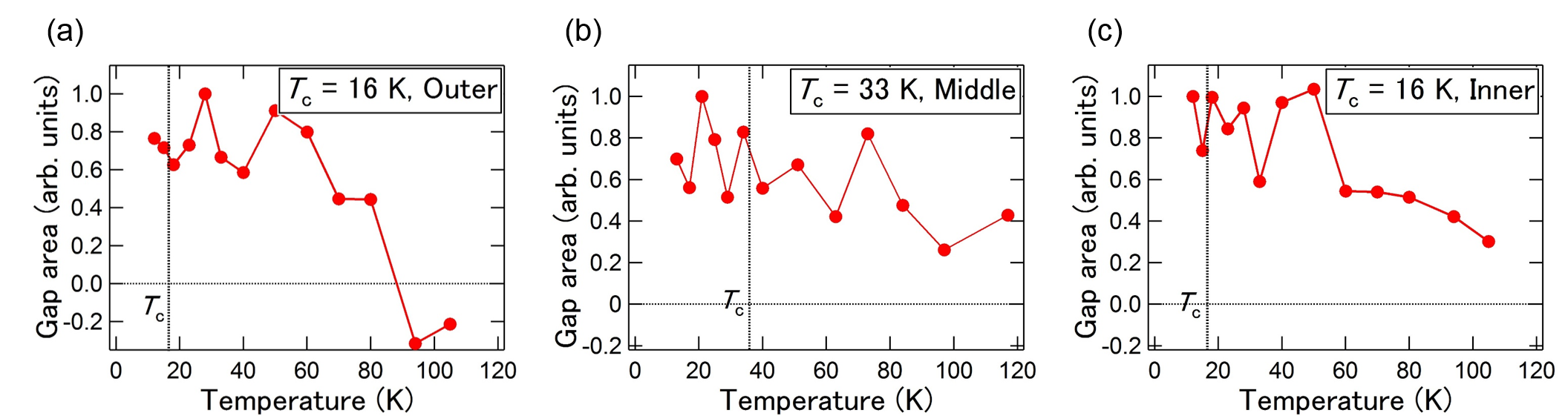}
\caption{
Gap area obtained from the linear-background-subtracted EDCs (Fig. S \ref{fig:S4})
}\label{fig:S5}
\end{figure*}

\begin{figure*}
\includegraphics[clip,width=14cm]{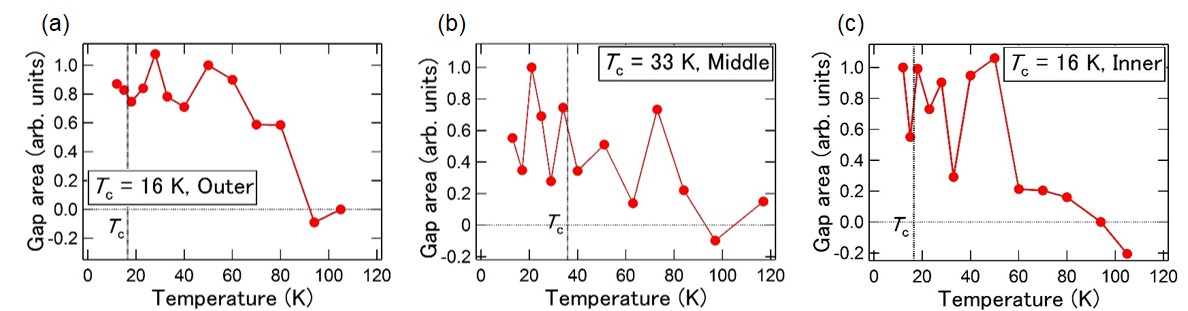}
\caption{
Normalized gap area obtained from the linear-background-subtracted EDCs (Fig. S \ref{fig:S4})
}\label{fig:S6}
\end{figure*}

\end{document}